\documentclass[fleqn,usenatbib]{mnras}

\usepackage{newtxtext,newtxmath}

\usepackage[T1]{fontenc}

\DeclareRobustCommand{\VAN}[3]{#2}
\let\VANthebibliography\thebibliography
\def\thebibliography{\DeclareRobustCommand{\VAN}[3]{##3}\VANthebibliography}

\usepackage{graphicx}	
\usepackage{amsmath}	
\usepackage{stfloats}
\usepackage{xcolor}

\title[B-mode inference with realistic foregrounds]{Single frequency CMB B-mode inference with realistic foregrounds from a single training image}

\author[N. Jeffrey et al.]{Niall Jeffrey,$^{1, 2}$\thanks{E-mail: niall.jeffrey@phys.ens.fr}
François Boulanger,$^{1}$
Benjamin D. Wandelt,$^{3,4}$ 
Bruno Regaldo-Saint Blancard,$^{1, 5}$ \and
Erwan Allys,$^{1}$
François Levrier$^{1}$\\
$^{1}$Laboratoire de Physique de l'École Normale Supérieure, ENS, Université PSL, CNRS, Sorbonne Université, Université de Paris, 75005 Paris, France \\
$^{2}$Department of Physics \& Astronomy, University College London, Gower Street, London, UK \\
$^{3}$Institut d'Astrophysique de Paris (IAP), UMR 7095, CNRS, Sorbonne Universit\'e, France \\
$^{4}$Center for Computational Astrophysics, Flatiron Institute, 162 5th Avenue, New York, USA \\
$^{5}$Observatoire de Paris,PSL University, Sorbonne Université, LERMA, 75014 Paris, France
}

\date{Accepted 2021.}

\pubyear{2021}

\begin{document}
\label{firstpage}
\pagerange{\pageref{firstpage}--\pageref{lastpage}}
\maketitle

\begin{abstract}
With a single training image and using wavelet phase harmonic augmentation, we present polarized Cosmic Microwave Background (CMB) foreground marginalization in a high-dimensional \textit{likelihood-free} (Bayesian) framework. We demonstrate robust foreground removal using only a single frequency of simulated data for a BICEP-like sky patch. Using Moment Networks we estimate the pixel-level posterior probability for the underlying $\{E,B\}$ signal and validate the statistical model with a quantile-type test using the estimated marginal posterior moments. The Moment Networks use a hierarchy of U-Net convolutional neural networks. This work validates such an approach in the most difficult limiting case: pixel-level, noise-free, highly non-Gaussian dust foregrounds with a single training image at a single frequency. For a real CMB experiment, a small number of representative sky patches would provide the training data required for full cosmological inference. These results enable robust \textit{likelihood-free}, simulation-based parameter and model inference for primordial B-mode detection using observed CMB polarization data. 
\end{abstract}

\begin{keywords}
cosmology: cosmic background radiation  -- methods: statistical
\end{keywords}


\section{Introduction} \label{sec:intro} \vspace{-0.1cm}
\enlargethispage*{0.4cm}
Polarized galactic foregrounds are the most significant challenge for the detection of the polarized cosmic microwave background (CMB) B-mode signal of primordial gravitational waves. Such a detection would provide a direct probe of cosmic inflation in the early Universe.

{\color{black} Typically, the cosmological signal is recovered using multi-frequency observations, for which there have been numerous recent developments in the context of Planck and ground-based experiments \citep[][]{planck_cleaning, darwish_act}. The sensitivity expected from the next generation of experiments poses further challenges \citep{simons, cmb_s4}.}

A significant hindrance is the difficulty either numerically simulating the polarized galactic dust emission or constructing analytic models. In particular, if it were possible to generate synthetic realizations of the dust polarization signal, the task could be remarkably simplified. With the ability to forward model by generating mock data, the use of simulation-based methods to perform \textit{likelihood-free inference} of unknown parameters is relatively straightforward.

Generating a significant number of polarized foreground signal realizations has been so far impossible. Numerical simulations are computationally costly and fraught with difficulties. Furthermore, low-noise and high-precision observational data is also limited. 

In this Letter, we show that with only a single training image of polarized dust emission we can use \textit{wavelet phase harmonic synthesis} to generate new representative samples. These samples form a training set with which we can perform likelihood-free inference. We test the high-dimensional posterior probability space of pixel values directly, which proves to be a powerful general validation of this combined synthesis and likelihood-free approach.

Using only a single frequency (143 GHz) we use a hierarchy of convolutional neural networks to form a Moment Network \citep{jeffrey2020solving} to estimate posterior mean and marginal variance per pixel of the extra-galactic CMB signal given the observed polarization data. \textit{Summary of our approach}: \begin{itemize}
\item \noindent  Simulate a single polarized foreground data patch.
\item \noindent  Synthesize fast realizations using  wavelet phase harmonics to form a large augmented set of foreground maps.
\item \noindent  Draw cosmological parameters from a prior probability distribution and generate mock CMB signals.
\item \noindent  Draw foreground amplitude parameters from a prior probability distribution to match a BICEP-like sky at a single 143 GHz frequency.
\item \noindent  Train Moment Network to estimate the marginal posterior mean and variance for the CMB signal for each pixel. 
\item \noindent  Validate the per-pixel posterior estimates with a quantile-type test using new \textit{unseen} simulated (not synthesized) foreground data.
\end{itemize} \vspace{-0.01cm}
By validating the pixel-by-pixel statistical model, we have demonstrated that this approach is robust for parameter inference and, by extension, tensor-to-scalar $r$ detection.

For application to real data, the inference task will be in many ways easier. Here we have chosen the set-up to be particularly difficult to demonstrate the strengths of the method: noise-free, a single training image, and a single frequency. Measurement noise will Gaussianize the data, so would limit our demonstration of highly non-Gaussian dust foreground removal. 

\cite{Aylor_2020} previously demonstrated an adversarial networks, using $>10^3$ cleaned (Planck) training images, for total intensity foreground cleaning. Using the results of this Letter, we can now use very few representative foreground data patches for polarization (B-mode) analysis as part of our new Bayesian validated likelihood-free approach.
\enlargethispage*{0.5cm}
\vspace{-0.4cm}
\section{Statistical model}
\enlargethispage*{0.3cm}
\vspace{-0.1cm}
\subsection{Likelihood-free inference}
\vspace{-0.1cm}
\paragraph{Overview:}  In Bayesian parameter inference of unknown parameters $\boldsymbol{\theta}$, we aim to evaluate the posterior probability distribution
\begin{equation} \label{eq:bayes}
    p(\boldsymbol{\theta} | \mathbf{d}_{obs} , \mathcal{M}) = \frac{p(\mathbf{d}_{obs} | \boldsymbol{\theta}, \mathcal{M}) \  p(\boldsymbol{\theta} | \mathcal{M})}{ p(\mathbf{d}_{obs} | \mathcal{M})}
\end{equation}
\noindent for some statistical model $\mathcal{M}$ given the observed data (or summary statistics of the observed data) $\mathbf{d}_{obs}$ (see~\citealt{jaynes07} for details).

\enlargethispage*{0.2cm}
In a typical analysis, the likelihood $\mathcal{L}(\boldsymbol{\theta}) = p(\mathbf{d}_{obs} | \boldsymbol{\theta}, \mathcal{M})$ is assumed or modelled analytically. In \textit{likelihood-free inference} (also known as simulation-based inference) the likelihood is not assumed. Instead, the sampling distribution of the data $p(\mathbf{d} | \boldsymbol{\theta}, \mathcal{M})$ as a function of the unknown parameters is estimated from forward modelled data.

With density estimation likelihood-free inference \citep{Papamakarios_lfi, delfi1, delfi2}, the inference task is posed as a density estimation problem. The simulated mock data $\mathbf{d}$ realizations and their respective parameters $\boldsymbol{\theta}$ form a cloud of points in $\{ \mathbf{d}, \ \boldsymbol{\theta} \}$ space. In this space we could estimate the following distributions: the joint $p( \mathbf{d}, \boldsymbol{\theta}, \mathcal{M})$, the conditional $p(  \boldsymbol{\theta} | \mathbf{d}, \mathcal{M})$, or the conditional $p( \mathbf{d} | \boldsymbol{\theta}, \mathcal{M})$ known as the sampling distribution (which becomes likelihood if evaluated for observed data $\mathbf{d}_{obs}$). 

Provided mock data  $\boldsymbol{d}$ can be generated, with each realization having an associated $\boldsymbol{\theta}$ label, the problem of inference given the actual observed data is relatively straightforward. Using this density estimation approach, a number of cosmological analysis have now been carried out~\citep[e.g.][]{Brehmer_2019, des_lfi, ramanah2020dynamical, 2lemos_lfi} that infer model parameters $\boldsymbol{\theta}$ without needing to assume or approximate a closed-form likelihood.

In this work, thanks to wavelet phase harmonic synthesis (section~\ref{sec:wph}), we now have the means to generate fast realizations of dust foregrounds for polarization CMB analysis (given our single training image). However, rather than using likelihood-free inference to infer only a set of cosmological parameters, here we choose to demonstrate a pixel-by-pixel inference at the level of the map as a powerful test of this approach.

\vspace{-0.3cm}
\paragraph{Moment Networks:} These are a simple hierarchy of fast neural regression models that compute increasing moments of any lower-dimensional marginal posterior density~\citep{jeffrey2020solving}.

For a pixel-level inference, each unknown pixel value of the underlying CMB signal $\mathbf{s}$ is a parameter to be inferred. The probability space of all pixels in a given $B$ signal map  $\mathbf{s}_B$ has a dimension of $\mathcal{D} = 256\times256>6.5\times10^4$. Evaluation of the full joint posterior probability at such a dimensionality $\mathcal{D}$ is intractable.

Moment Networks side-step the problem of estimating the posterior density. Particularly useful for high-dimensional parameter spaces, they directly estimate moments of the marginal posterior distribution of single parameters or subsets of parameters. For example, we could estimate the mean and variance of the marginal posterior probability for each pixel of some unknown signal $\boldsymbol{\theta} = \mathbf{s}$,
\begin{equation} \label{eq:marg}
p({s}_\alpha | \mathbf{d}_{obs}, \mathcal{M} ) = \int  p({s}_\alpha ,\boldsymbol{s}' | \boldsymbol{d}_{obs}, \mathcal{M} ) \  \mathrm{d}  \boldsymbol{s}' \ \ ,
\end{equation}
where ${s}_\alpha$ is any given pixel (element of the signal vector) and $\boldsymbol{s}'$ are all other pixels. We could also estimate moments (e.g. covariance) of two-dimensional marginal posteriors for pairs of parameters:
\begin{equation} \label{eq:marg2d}
p({s}_\alpha, {s}_\beta | \mathbf{d}_{obs}, \mathcal{M} ) = \int   p({s}_\alpha, {s}_\beta ,\boldsymbol{s}' | \boldsymbol{d}_{obs}, \mathcal{M} ) \  \mathrm{d}  \boldsymbol{s}' \ \ .
\end{equation}
We require that the unknown parameters are drawn from prior distribution $\mathbf{s}_i \sim p(\mathbf{s} | \mathcal{M})$ and, through the forward model, the associated data are drawn from the sampling distribution  $\mathbf{d}_i \sim p(\mathbf{d} | \mathbf{s}_i, \mathcal{M})$. The first layer of the hierarchy in the Moment Network finds some function $\mathcal{F}(\boldsymbol{d})$ of our data that minimizes a squared loss over the distribution of possible training examples $\{ \mathbf{d}_i,  \mathbf{s}_i \}$,
\begin{equation}
    J_0 = \int || \boldsymbol{s} - \mathcal{F}(\boldsymbol{d}) ||^2 p(\boldsymbol{d},  \boldsymbol{s}  ) \ \mathrm{d}\boldsymbol{d} \  \mathrm{d}\boldsymbol{s} \ \ ,
\end{equation}
\noindent then $\mathcal{F}$, which is a neural network, evaluated for the observed data is the mean of the posterior distribution $\mathcal{F}(\boldsymbol{d}_{obs}) = \langle \boldsymbol{s} \rangle_{p(\boldsymbol{s} | \boldsymbol{d}_{obs})}$.

Moment Networks allow the use of far simpler neural network architectures, which reduces training failure risks and improves inference speed, and have recently been applied to Cosmic Void inference~\citep{void_mn}.

In this work, at the next level of the hierarchy, the function $\mathcal{G}$ minimizes
\begin{equation}
    J_1 = \int || (\boldsymbol{s} - \mathcal{F}_{\mathrm{fixed}}(\boldsymbol{d}))^2 - \mathcal{G}(\boldsymbol{d}) ||^2 p(\boldsymbol{d},  \boldsymbol{s}) \  \mathrm{d}\boldsymbol{d} \  \mathrm{d}\boldsymbol{s} \ \ ,
\end{equation}
for fixed, already trained $\mathcal{F}$. If $J_1$ is minimized over the training data (drawn from the correct prior and forward model) then $\mathcal{G}(\boldsymbol{d}_{obs})$ gives the posterior variance for the marginalized posterior of the signal for each pixel. This result is exact and independent of the true underlying posterior or prior distributions~\citep{jaynes07,adler}. 

In this work, we train the first network $\mathcal{F}$ of the two-layer hierarchy to return $ \boldsymbol{\mu}_B=\langle \boldsymbol{s}_B \rangle_{p(\boldsymbol{s} | \boldsymbol{d}_{obs})}$, which is the mean of the marginal posterior of the signal $\mathbf{s}_B$ for each pixel given the data \{$\mathbf{d}_E$,$\mathbf{d}_B$\}:
\begin{equation}
\mathcal{F}(\{\boldsymbol{d}_{E},\boldsymbol{d}_{B}\}) = \int  \boldsymbol{s}_B \  p(\boldsymbol{s}_B| \boldsymbol{d}_{E},\boldsymbol{d}_{B}, \mathcal{M} ) \  \mathrm{d}  \boldsymbol{s}_B \ \ . \
\end{equation}
The second network $\mathcal{G}$ gives the variance $\sigma_B^2 = \langle \boldsymbol{s}_B^2 \rangle_{p(\boldsymbol{s} | \boldsymbol{d}_{obs})} - \boldsymbol{\mu}_B^2 $ of the marginal posterior distribution for each pixel of $\mathbf{s}_B$  given the data \{$\mathbf{d}_E$,$\mathbf{d}_B$\}:
\begin{equation}
\mathcal{G}(\{\boldsymbol{d}_{E},\boldsymbol{d}_{B}\}) = \int  \big( \boldsymbol{s}_B^2 -  \boldsymbol{\mu}_B^2  \big)\  p(\boldsymbol{s}_B| \boldsymbol{d}_{E},\boldsymbol{d}_{B}, \mathcal{M} ) \  \mathrm{d}  \boldsymbol{s}_B \ \ . \
\end{equation}
In this work we show the result for $\mathbf{s}_B$ here as it is the most significantly contaminated by polarized dust foregrounds. Both stages of the Moment Network use a UNet convolutional neural network from the \href{https://github.com/NiallJeffrey/DeepMass}{\texttt{DeepMass}}\footnote{\url{https://github.com/NiallJeffrey/DeepMass}} package~\citep{deepmass} -- see \autoref{sec:results}.
\vspace{-0.23cm}
\subsection{CMB data model} \label{sec:cmb}
\vspace{-0.15cm}
\enlargethispage*{0.1cm}
We select a flat $\Lambda$-Cold Dark Matter ($\Lambda$CDM) cosmological model for this analysis. This choice of model is not a requirement for this approach; one could choose any cosmological model $\mathcal{M}$ provided it is possible to generate mock observables where the unknown cosmological parameters $\boldsymbol{\Theta}_C$ are drawn from some prior probability distribution $p(\boldsymbol{\Theta}_C| \mathcal{M})$.

The parameters used for the CMB signal generation must be drawn from the assumed prior $\boldsymbol{\Theta}_{C,i} \sim p(\boldsymbol{\Theta}_C | \mathcal{M})$, so that the Moment Network output matches the expected properties of the posterior probability distribution. In this work we choose to use the~\cite{planck_cosmo}  \texttt{base\_plikHM\_TTTEEE\_lowl\_lowE}\footnote{\url{https://pla.esac.esa.int/}} parameter posterior distribution as our prior.

Our sampled cosmological parameters are: the Hubble parameter $H_0$, the baryon density $\Omega_b$, the cold dark matter density $\Omega_c$, the Reionization optical depth $\tau$, primordial
amplitude $A_s$, scalar spectral index $n_s$. We fix the tensor-to-scalar ratio parameter $r = 0$, but this could be easily allowed to vary over a wide prior range for such an analysis. We thin the Markov chain Monte Carlo Planck chains uniformly to sample approximately $3000$ sets of parameters $\boldsymbol{\Theta}_{C,i}$.

For each set of parameters drawn from the prior, we calculate the theoretical CMB polarization $EE$ and $BB$ power spectrum $C_\ell(\boldsymbol{\Theta}_{C,i})$ using the \textsc{camb} software~\citep{camb1, camb2}. 
For each $C_\ell(\boldsymbol{\Theta}_{C,i})$ sample, we generate a full-sky Gaussian realization of the $E$ and $B$ signals with HEALPix~\citep{healpix1}. These Gaussian signal realizations do not include higher-order N-point corrections caused by any primordial non-Gaussianity or gravitational lensing, and therefore would be insufficient for a full delensing analysis, though such signals would not be difficult to simulate if required~\citep[][]{millea}. From each full-sky HEALPix map (resolution \texttt{nside}=1024), we extract 18 non-overlapping patches using gnomonic projection for both $s_E$ and $s_B$. Each patch is $20^{\circ} \times 20^{\circ}$ to match a BICEP-like survey area~\citep{bicep} with 256$\times$256 pixels. Each $s_E$ and $s_B$ map is convolved with a Gaussian beam smoothing with the size of a single pixel width ($\sigma \approx 4.69$ arcmin). In total, we generate 52,224 sets of \{$s_E$, $s_B$\} maps.
\begin{figure*}
\vspace*{-0.35cm}
\centering
\hspace*{-0.1cm}
\includegraphics[width=1.0\textwidth]{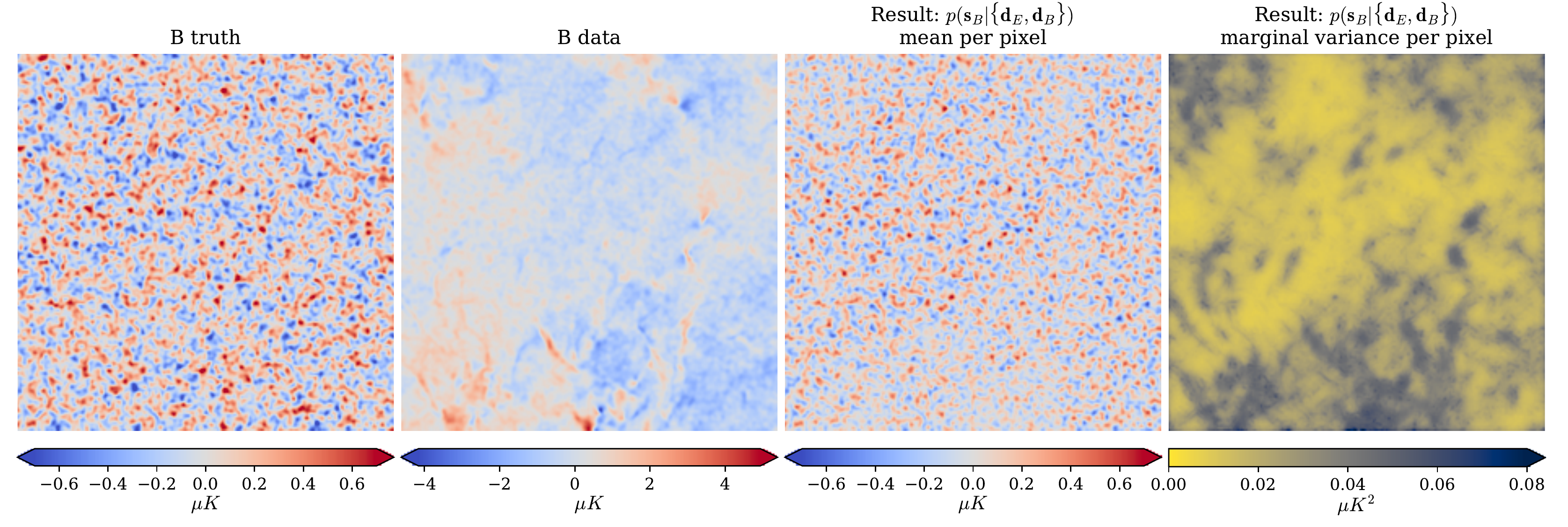}
\vspace*{-0.8cm}
\caption{\label{fig:maps} The two left panels show the simulated clean signal $\mathbf{s}_B$ and the foreground-contaminated data $\mathbf{d}_B$ (validation data A - \autoref{sec:validation}). The centre right panel shows the mean of the marginal posterior probability per pixel $\mathcal{F}( \mathbf{d}_{E}, \mathbf{d}_{B})$ and the far right shows the variance of the marginal posterior per pixel $\mathcal{G}( \mathbf{d}_{E}, \mathbf{d}_{B})$. This CMB signal has been inferred (the posterior probability estimated) using only a single frequency and a single training image. Patches of reduced power in the pixel posterior mean are not artefacts; the mean is expected to move closer to 0 $\mu K$ when the posterior variance is higher.}
\vspace*{-0.5cm}
\end{figure*} 

In reality, the CMB polarization signal is measured in terms of the \{$Q,U$\} Stokes fields, rather than the pseudo-scalar\{$E,B$\} fields~\citep{qu_cmb}, which generally introduces artefacts when transforming between these representations for masked or non-periodic data. However, this is already a solved problem in a single frequency, unlike polarization dust foreground marginalization. Again, though this $E$-$B$-leakage effect should be included for analyses with actual observational data, adding this effect here would diminish our clear demonstration of non-Gaussian foreground marginalization.

The left panel of \autoref{fig:maps} shows an example $B$ signal simulation. The centre-left panel shows the same map with simulated foregrounds significantly obscuring the signal. 
\vspace{-0.25cm}
\subsection{Foreground modelling} \label{sec:foregrounds}
\vspace{-0.1cm}
\paragraph{Wavelet phase harmonic synthesis:} \label{sec:wph}
Using a single simulation of the polarized dust foregrounds $\mathbf{f}^*_Q$, $\mathbf{f}^*_U$, we synthesize many further realizations using wavelet phase harmonic synthesis \{$\mathbf{f}_Q, \mathbf{f}_U$\}. The original simulated foreground image $\mathbf{f}^*_Q$, $\mathbf{f}^*_U$
is of a polarization signal as produced by the thermal emission of dust
in the diffuse ISM. It is built from the same magnetohydrodynamic (MHD) simulation as described in~\cite{mhd}.

Wavelet phase harmonic statistics are descriptions of signals that have been designed taking advantage of comparisons between convolutional neural networks and non-linear harmonic analysis operations. These statistics form models that can be used to generate new realizations of a given signal.

We can define the wavelet phase harmonic coefficients $\boldsymbol{\alpha}$ of a foreground map in relation to a wavelet phase harmonic operator, where $\boldsymbol{\alpha}= \phi(\mathbf{f}^*)$. These statistics are designed to characterize the coherent structures that appear in non-Gaussian random fields, by quantifying the phase alignment between different scales~\citep{Zhang2019, Mallat2020}. For cosmic-web density fields, \cite{Allys2020} demonstrated that these statistics include most of the information captured by various high-order statistics. 

For a detailed mathematical description, we refer the reader to Appendix A of~\cite{bruno_denoising}, in which these statistics were measured for noisy dust polarization Planck data. However, for an overall understanding of wavelet phase harmonic statistics, the reader may consider that these statistics emulate similar properties to information capture in a convolutional neural network. Unlike neural networks, these statistics require no training, so it is possible to estimate the phase harmonic coefficients from our single training image  $\boldsymbol{\alpha}^*= \phi(\mathbf{f}^*)$. New synthesized $\mathbf{f}$ maps are generated by minimizing a loss,
\begin{equation}
\label{eq:loss}
    W_\alpha(\mathbf{f}) = || \phi(\mathbf{f}) - \boldsymbol{\alpha}^* ||^2,
\end{equation}
with $\mathbf{f}$ initialized as Gaussian white noise. Using differentiable $\phi$ operators, the \href{https://github.com/bregaldo/pywph}{\texttt{pyWPH}}\footnote{\url{https://github.com/bregaldo/pywph}} code uses gradient-based optimization.

As we are working in $Q,U$ space to synthesize, we can apply the operator directly on the complex field (as in~\citealt{bruno_denoising}). Here we extend this to include the individual components, giving a new loss: $W'_\alpha = W_\alpha(\mathbf{f}_Q) + W_\alpha(\mathbf{f}_U) + W_\alpha(\mathbf{f}_Q + i\mathbf{f}_U )$.

A further development of phase harmonic synthesis in this work is the use of differentiable histograms to match the Kullback–Leibler divergence of the 1-point distributions of the synthesized $\mathbf{f}_Q$ and $\mathbf{f}_U$ fields with the target (see Appendix B - Supporting Materials).

These techniques combine to give a realistic new set of synthesized polarized foregrounds that match the single initial training simulation. The robustness of the synthesized maps for this problem is tested in the following section.

\begin{figure*}
\vspace*{-0.5cm}
\centering
\hspace*{-0.6cm}
\includegraphics[width=1.04\textwidth]{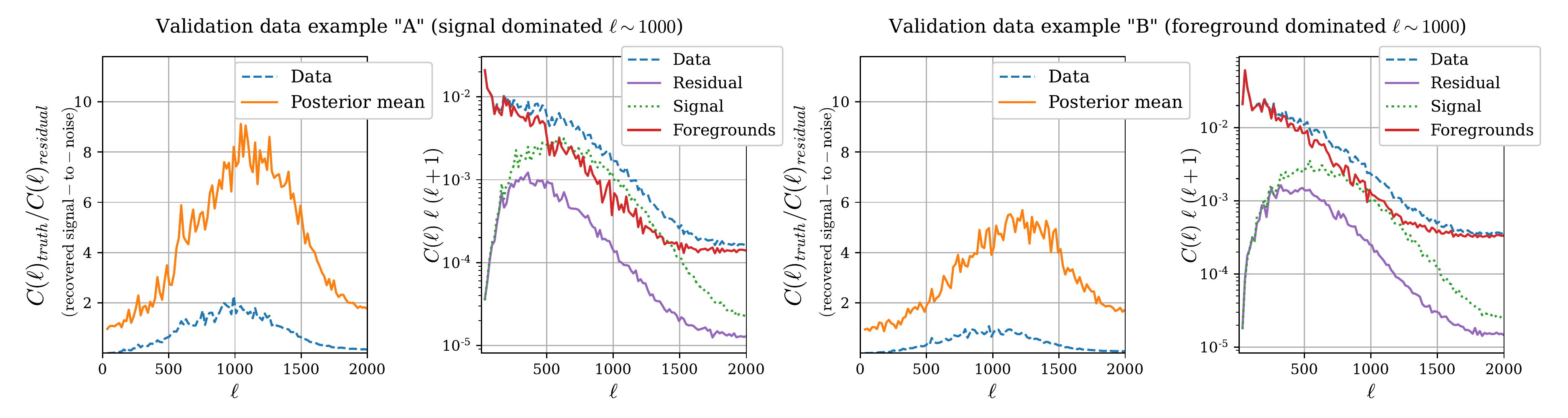}
\vspace{-0.7cm} 
\caption{\label{fig:joint}Each pair of panels on the left and right correspond to the validation data examples ``A'' and ``B'' respectively. The right panel of each pair shows the power spectra of data components and residuals. The left panel shows the ratio of the true power to the power of the residuals, between the posterior mean and the truth (\textit{orange solid line}) or between the data and the truth (\textit{blue dashed line}); ${C(\ell)_{truth}}/{C(\ell)_{residual}}$ can be interpreted as signal-to-noise (see~\autoref{sec:cleaning} for discussion). If interpreted as a foreground-cleaned map, the posterior mean per pixel shows a significant improvement over the data, as would be expected.}
\vspace{-0.4cm} 
\end{figure*} 

\vspace{-0.25cm}
\paragraph{Foreground amplitude prior:} \label{sec:foreground_amplitude}
We match the synthesized foreground maps, which have 512$\times$512 pixels, to a $40^\circ \times 40^{\circ}$ area. For each periodic synthesized map, we transform from $Q,U$ to $E,B$. We then select sub-patches of 256$\times$256 pixels to break the periodic boundary conditions and to match the survey size of the simulated CMB signal maps. From 102 full 512$\times$512 synthesized foreground maps, we use data augmentation methods of image flips, translations and rotations (applied equally to $\mathbf{f}_E$ and $\mathbf{f}_B$) to generate a set of 52,224 foreground maps of size 256$\times$256.

These synthesized foreground maps are matched to realistic foreground properties for a BICEP-like experiment. We model the polarized dust foreground amplitude of the BICEP field using the Planck analysis values and uncertainties. We first assume that the large scale foregrounds are known well enough that we can subtract the mean foreground signal so that $\langle \mathbf{f}_E \rangle = \langle \mathbf{f}_B \rangle = 0$ for each map. 
The foreground amplitude parameters $A_{E}$ and $A_{B}$ are defined with respect to a power-law spectrum:
\begin{equation} \label{eq:foreground_power}
C_\ell^{XX} = A_{X} \ (\ell/80)^{-0.42} \ (2 \pi ) \ / \ ( \ell \ (\ell+1) ) \\ .
\end{equation}
The Planck analysis values and uncertainties for each $A_{X}$ contribute to our prior $p(A_{E}, A_{B})$. As a conservative least-informative approach, we assume independent Gaussians: $ p(A_{E}) = \mathcal{N}(0.358, 0.0917^2)$  \& $\mathcal{N}(0.172, 0.0440^2)$, where the model and values are derived from \cite{bicep_planck} section 3.2.

To match these amplitudes defined on the celestial sphere to the pixel variance of our foregrounds patches, we generate 1000 full-sky Gaussian realizations of foreground maps with power spectrum parameters drawn from the above prior. From these full-sky realizations, we create patches with the correct angular size and resolution (these realizations are then discarded). This Monte Carlo procedure transforms the prior distribution with respect to the power spectrum amplitude parameters $p(A_{E}, A_{B})$ to a prior  on the pixel standard deviation of the patches $p(\sigma_{E}, \sigma_{B})$. The synthesized foreground maps \{ $\mathbf{f}_{E,i}$, $\mathbf{f}_{B,i}$ \} are rescaled to match the sampled amplitudes.
\enlargethispage*{0.3cm}
\vspace{-0.3cm}
\section{Results} \label{sec:results}
\vspace{-0.2cm}
\subsection{Marginal posterior moments} \label{sec:results1}
\vspace{-0.1cm}
\enlargethispage*{0.1cm}
The previous section described the statistical model that generated mock data samples. For each $i$ sample, we have: foreground contaminated data maps ($\mathbf{d}_{E,i}$, $\mathbf{d}_{B,i}$), clean signal maps ($\mathbf{s}_{E,i}$, $\mathbf{s}_{B,i}$), associated cosmological and foreground parameters (${A}_{E,i}$,${A}_{B,i}$, $\mathbf{\Theta}_{C,i}$).

In this work, we target the posterior probability distribution of the pixel values of the $B$-mode signal given the data $p(\mathbf{s}_B | \mathbf{d}_{E}, \mathbf{d}_{B})$ with all other parameters marginalized away. We choose the $B$-mode signal for this demonstration as the signal power is much weaker than $E$-mode (by a factor of 300) so is most obscured by polarized dust foregrounds. By validating the posterior probability at pixel-level, this is a powerful demonstration of the robustness of the synthesized forward model and the likelihood-free inference.

For both stages of the Moment Network, the posterior mean and variance networks, $\mathcal{F}( \mathbf{d}_{E}, \mathbf{d}_{B})$ and $\mathcal{G}( \mathbf{d}_{E}, \mathbf{d}_{B})$, we train the UNet convolutional neural network from \href{https://github.com/NiallJeffrey/DeepMass}{\texttt{DeepMass}} for 100 epochs with varied batch size and learning rates (Convergence is achieved at epoch 50). From the synthesized training samples we retain 5120 as a test set to confirm that over-fitting does not occur. Moment Networks are fast and inexpensive, taking 94 seconds per epoch to train with an \texttt{Nvidia V100 32GB} and evaluation of moments taking $\approx1$ms per new data map. 

To validate the inference we do not use further synthesized images, but we take a new unseen simulation. We use four non-overlapping foreground patches from an approximately-independent snapshot of the same steady-state MHD simulation (see section 3.1 of ~\citealt{mhd} for details). The same steps of the forward model are applied to these new patches, including drawing the foreground amplitude from a prior for each map. Four new simulated CMB signals are added to generate four new data sets, which are labelled A, B, C, and D. 

\autoref{fig:maps} shows the results of the trained Moment Network applied to validation data A. The left panels show the clean target $\mathbf{s}_B$ and the foreground-contaminated data $\mathbf{d}_B$, though the posterior is conditioned on the $E$- and $B$-mode data. This figure is a demonstration of 5 of the 6 targets of the "Summary of our approach" described in the Introduction~(\autoref{sec:intro}).

Though naively the behaviour of the mean map appears to match what would be expected from a linear filter (e.g. Wiener filtering of noise), here, at a single frequency, the ability to distinguish between target signal and the statistically-homogeneous foregrounds is entirely due to non-Gaussianity. Appendix B (Supporting Materials) shows the same example with a Gaussian foreground model.
\begin{figure*}
\centering
\hspace*{-0.05in}
\includegraphics[width=.87\textwidth]{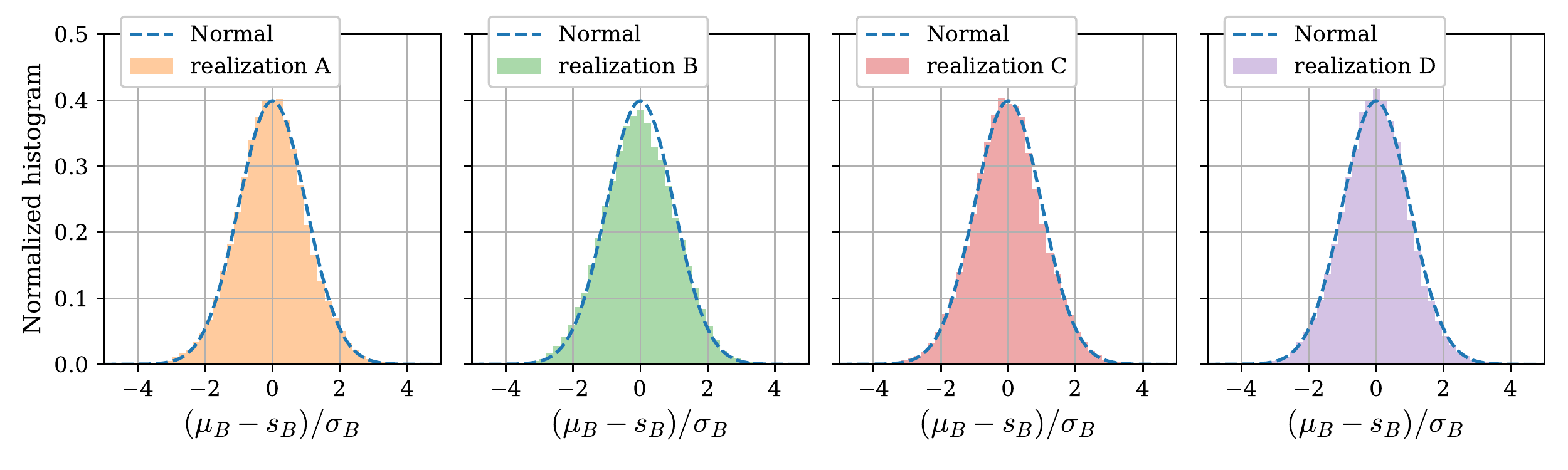}
\vspace{-0.5cm}
\caption{\label{fig:quantile} For each of the validation data maps (A, B, C, D), the distribution of the rescaled pixel residuals $(\boldsymbol{\mu}_B - \boldsymbol{s}_B)/\boldsymbol{\sigma}_B$. This distribution is expected to have mean zero and unit standard deviation; the apparent Gaussianity is not necessarily expected or required and, though surprising, shows that the marginal posterior distributions are Gaussian for this data model. This result validates both the synthesis procedure with a single training image and the high-dimensional likelihood-free inference with Moment Networks.} \vspace{-0.4cm}
\end{figure*} 
 \vspace{-0.3cm}
\subsection{Posterior mean as foreground cleaning} \label{sec:cleaning}
\enlargethispage*{0.2cm}
\vspace{-0.15cm}
The aim of this high-dimensional likelihood-free inference approach is to estimate properties of the pixel posterior distribution. By validating this procedure, we can be confident in the forward model and the inference scheme for cosmological parameters or models.

However, the mean of the pixel posterior distribution is expected to be close to the true pixel values. As such, we can interpret the centre right panel of \autoref{fig:maps} as a foreground-cleaned map. It is important to note that the mean of the posterior is not a sample from the posterior, so it does not have the properties of \textit{in-painting}; pixels with high uncertainty are expected to have values close to the global mean $0 \mu K$. This occurs when the foreground signal is particularly obscuring, which are accompanied by an appropriate increase in the marginal variance per pixel. We can test the accuracy of the recovered pixel values, provided we remember they are the mean values of a high-dimensional posterior probability $p(\mathbf{s}_B | \mathbf{d}_{E}, \mathbf{d}_{B})$. 

The two left panels of \autoref{fig:joint} show results from validation data A and the two right panels show result from validation data B. The right panel of each pair, shows the power spectra of the components and residuals. We can see that the signal power dominates the foreground at $\ell \approx 1000$ for validation data A but the foregrounds always dominate for validation data B. Though not shown, example C is neither foreground- nor signal-dominated and D is signal dominated.

The residuals are defined as the difference between the truth and a given map, either the original data $\boldsymbol{d}_B$ or mean posterior map $\boldsymbol{\mu}_B$. The left panels of each pair show the power spectrum of the truth divided by the power spectrum of the residual; this quantity (truth divided by an effective error) can be interpreted as a signal-to-noise. We clearly see that if we take the mean per pixel as a cleaned map, we significantly reduce the relative error compared with the foreground contaminated data.
\vspace{-0.25cm}
\subsection{Posterior probability validation} \label{sec:validation}
\vspace{-0.2cm}
For each pixel value, we have a marginal posterior mean and variance. As this is simulated data and we have a large number of pixels, we can validate that the true values are distributed around the mean with the correct variance as predicted by our inference pipeline. In particular the rescaled pixel residuals $(\boldsymbol{\mu}_B - \boldsymbol{s}_B)/\boldsymbol{\sigma}_B$ are expected to have zero mean and unit variance.

\autoref{fig:quantile} shows the rescaled residual distributions for each of the validation data maps A, B, C, and D. The clear Gaussianity of the residuals is not generally expected, but indicates that the marginal posterior distributions are Gaussian for this data model. The high level of agreement between a unit Normal distribution and the rescaled residuals for each of the validation maps is a strong demonstration that the forward model and the high-dimensional likelihood-free inference is robust. This is the primarily validation of our approach.
\vspace{-0.4cm}
\section{Conclusions}
\enlargethispage*{0.9cm}
\vspace{-0.18cm}
In this Letter we have described two significant contributions to simulation-based inference of CMB polarization data. The first is the development of WPH synthesis to take a single training image of polarized dust foregrounds to create a large set of realistic training images for use in a forward model. The second is the application of high-dimensional likelihood-free methods, in particular Moment Networks, to target the posterior probability of the CMB signal pixel values given our observed data. Finally we validate both aspects by performing a quantile-type test using mock data derived from an unseen (not synthesized) simulation of dust foregrounds.

The quantile-type test validates our approach in the most difficult limiting case: pixel-level, noise-free, highly non-Gaussian dust foregrounds with a single training image at a single frequency. It would be straightforward to combine this single-frequency (morphological) approach with existing multi-frequency models. Future interesting developments would include astrophysical modelling of dust frequency decorrelation~\citep{decorrelation}, improvements in the accuracy and efficiency of WPH synthesis, and extensions of the likelihood-free inference methods. But now, with a few representative patches of the galactic microwave sky, either simulated or data-driven, we can overcome a significant challenge of dust emission for CMB polarization analysis in a likelihood-free framework.

\vspace{-0.5cm}
\section*{Data Availability}
\vspace{-0.15cm}
Reproducible code demonstrations and associated data are available on the DeepMass repository: \url{https://github.com/NiallJeffrey/DeepMass}
\vspace{-.3cm}
\enlargethispage*{0.65cm}

\section*{Acknowledgements}
\vspace{-0.15cm}
This research was supported by the Agence Nationale de la Recherche (Project BxB: ANR-17-CE31-0022).
\vspace{-0.15cm}



\label{lastpage}
\bibliographystyle{mnras}
\bibliography{example} 


%
\vspace{-0.3cm}
\section*{Supplementary material}
\vspace{-0.2cm}
\appendix
\section{Histogram Kullback-Leibler matching} \label{append:klmatching}

\begin{figure}
\hspace*{0.2cm}
\vspace*{-0.5cm}
\includegraphics[width=0.44\textwidth]{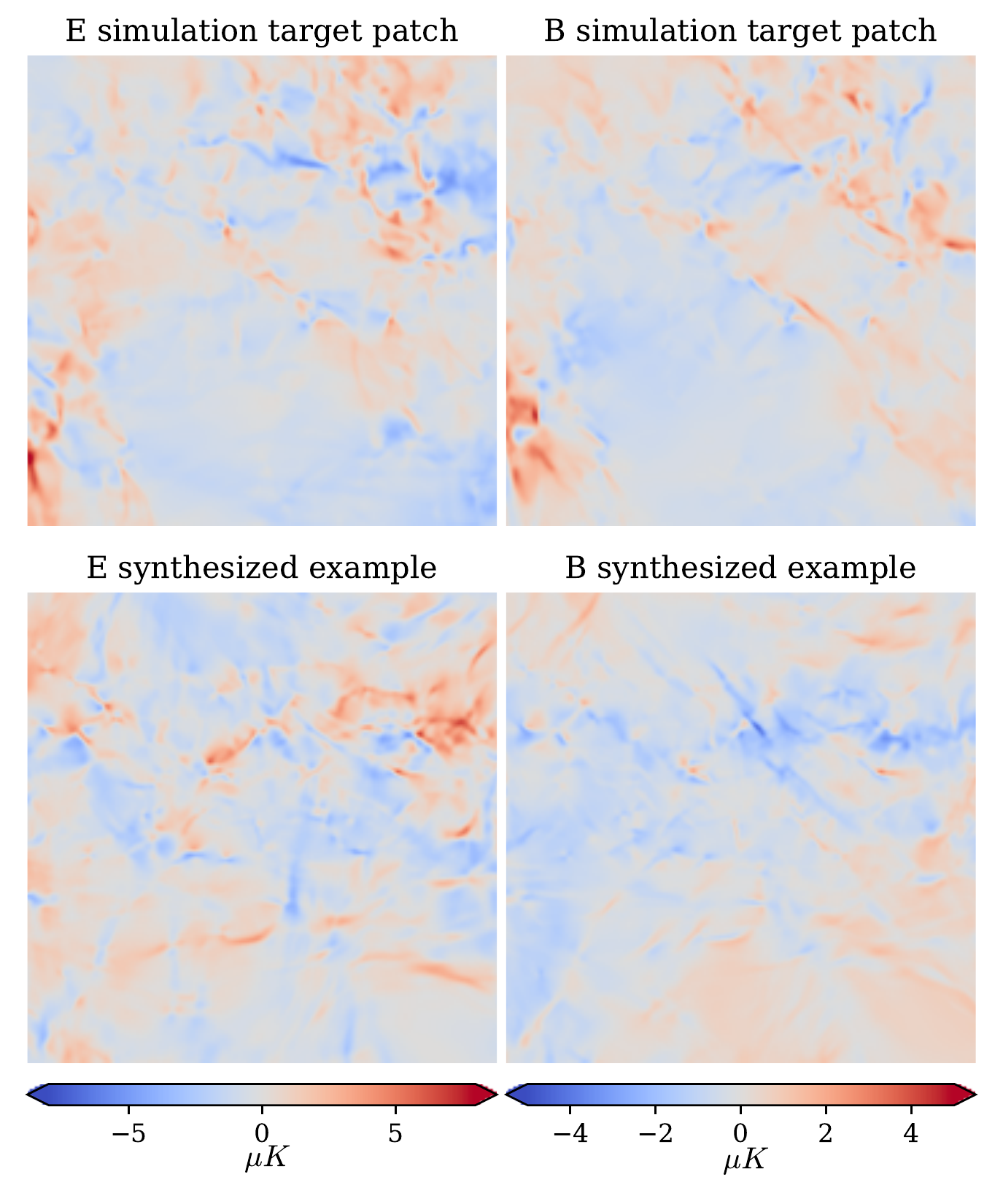}
\caption{\label{fig:synth_example} \textit{Upper row}: Simulated $\{E,B\}$ foregrounds as the target $\mathbf{f}^*$.  \textit{Lower row}: Resulting patch $\mathbf{f}$ of a typical synthesis as used in this work.}
\vspace*{-0.35cm}
\end{figure}

To ensure the 1-point distribution of a given synthesized signal $\mathbf{f}$ matches that of the target signal $\mathbf{f}^*$, we extend the loss function of the WPH optimization (\autoref{eq:loss}) to include a term to match the histogram of the synthesis to the target.  \autoref{fig:synth_example} shows a patch of the target simulation $\mathbf{f}^*$ and a resulting synthesized patch $\mathbf{f}$.

From the initial WPH loss $W$, we define a new loss $\mathcal{W} = W + \lambda J$, where the $J$ term encourages matched histograms. A histogram $\mathbf{q}$ of the synthesis $\mathbf{f}$ is a discrete estimate of a continuous density, where each element $q_n$ corresponds to a distinct histogram bin. To ensure that the density $\mathbf{q}$ matches the density $\boldsymbol{\rho}$ of the target signal $\mathbf{f}^*$, we minimize the \citealt{kullback1951} (KL) divergence,
\begin{equation} \label{eq:kl}
    J = \sum_n \rho_n \ln \rho_n - \rho_n \ln ( q_n + \epsilon) \ \ ,
\end{equation}
\noindent where the KL divergence is recovered in the limit $\epsilon \rightarrow 0$, but for which we have added a value of $\epsilon=10^{-8}$ for numerical stability. If the estimate of the KL divergence (also known as \textit{relative entropy}) is small, the two histograms are close. As we have a complex ($Q$,$U$) signal, we add a $\lambda J$ term for each component in the loss. The $\lambda$ value in the new loss $\mathcal{W}$ is a Lagrangian multiplier and is free to be chosen for this optimization problem. We use a value $\lambda = 5 \times 10^4$ that is typically much larger than the final WPH loss $W$, which ensures that the histograms are matched (with the WPH synthesis being constrained by this requirement).

Histograms are not typically differentiable, preventing gradient-based optimization of the KL term (\autoref{eq:kl}). We therefore use \textit{soft histograms} with overlapping bins that drop smoothly towards zero. Each pixel value is allocated to more than one bin, but with appropriate weight \citep[for details, see][]{forssen2001image}. We are therefore able to evaluate gradients of the loss with respect to pixel values, so can update the values to achieve the desired 1-point distribution.

\vspace*{-0.3cm}
\section{Naive Gaussian model} \label{append:naive_gauss}

\vspace*{-0.1cm}
As discussed in~\autoref{sec:results1}, the ability to perform a useful inference of the underlying signal $\mathbf{s}_B$ given the data $\{\mathbf{d}_E, \mathbf{d}_B\}$ relies on the non-Gaussianity of the foregrounds.

Here we demonstrate a simple statistical model where we naively assume that the dust foregrounds can be modelled as a Gaussian random field. We can assume $p(\mathbf{f}_B) = \mathcal{N}(0, \mathbf{C}_\mathbf{f})$ with known covariance $\mathbf{C}_\mathbf{f}$. We assume that the foregrounds are statistically isotropic and homogeneous (as for the non-Gaussian case in the main body of the paper), so $\mathbf{C}_\mathbf{f}$ is entirely defined by the power spectrum of $\mathbf{f}_B$. In this case, the posterior is that of a Wiener posterior with the foregrounds acting as noise~\citep{wandelt_gibbs}, which we can solve analytically (though for simplicity we apply a 3 pixel border mask).

\begin{figure}
\hspace*{.9cm}
\vspace*{-0.3cm}
\includegraphics[width=0.36\textwidth]{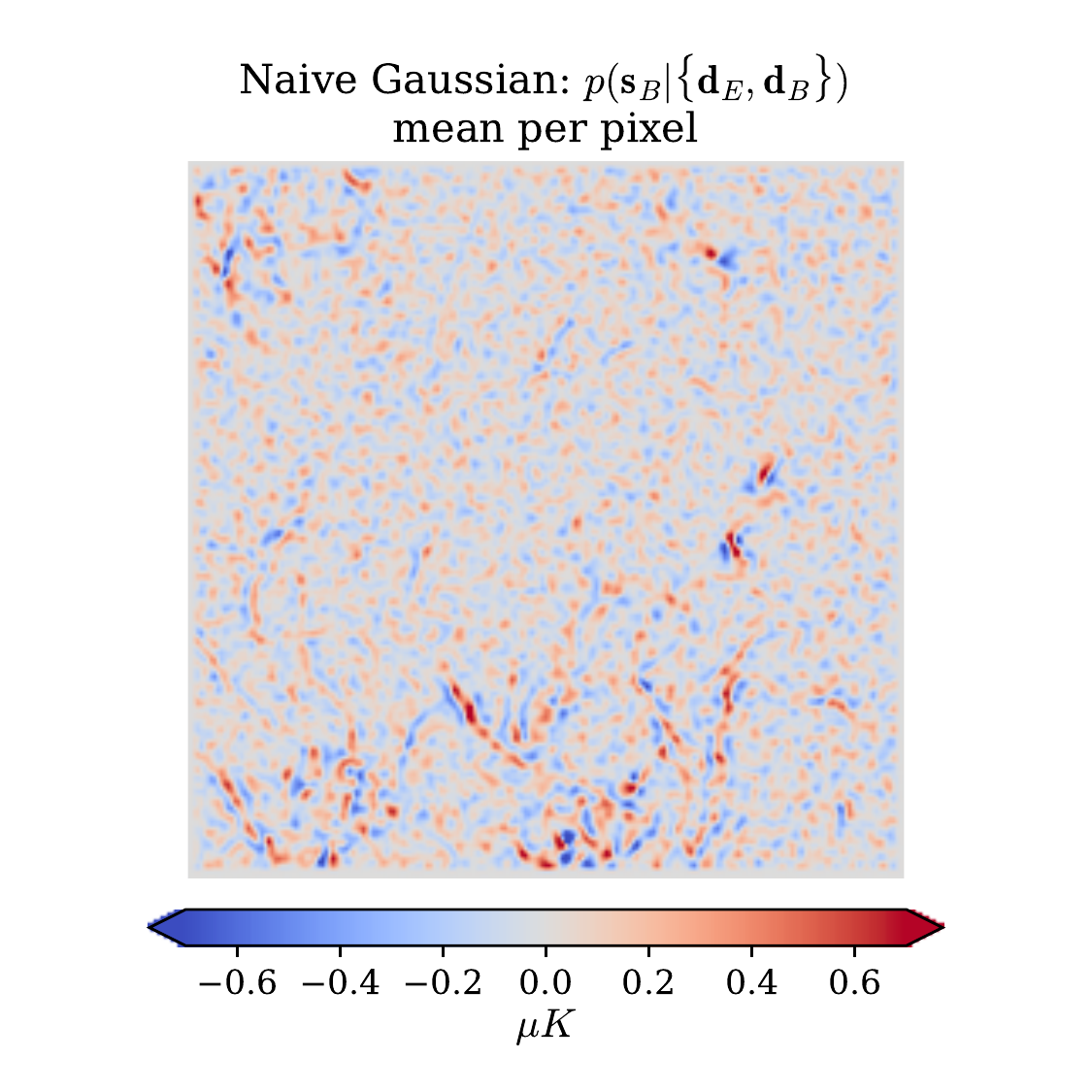}
\vspace*{-0.2cm}
\caption{\label{fig:naive_gauss} Validation data example ``A'': posterior mean using an analytic Gaussian statistical model, rather than a non-Gaussian forward model using likelihood-free inference.}
\vspace*{-0.3cm}
\end{figure}

The mean of the resulting posterior distribution, given validation data A, is shown in \autoref{fig:naive_gauss}, where we have assumed the correct power spectra of the signal and noise. As the signal and foreground components are (\textit{a priori}) homogeneous, the effective filtering is also homogeneous. Where the non-Gaussian foregrounds have particularly high-power, multiple artefacts become apparent in the mean posterior map. These errors are not represented in the variance of the marginal posterior per pixel. Though not shown, the variance per pixel is uniform across the field due to the homogeneity of the posterior distribution. The quantile-tests would therefore fail completely.

This shows that: (1) the non-Gaussianity of the foregrounds are the key to this type of component separation, and (2) naive Gaussian models would result in errors in the pixel posterior (unlike our likelihood-free Moment Network approach).

\enlargethispage*{0.5cm}


\end{document}